\documentclass[letterpaper, 11pt,conference]{IEEEtran} 
\IEEEoverridecommandlockouts                              % This command is only
\usepackage{graphicx, tabularx} % for pdf, bitmapped graphics files
\usepackage{bm}

\usepackage[superscript,biblabel]{cite}
\usepackage{booktabs}

\usepackage{titlesec}
\titlelabel{\thetitle.\quad}

\renewcommand{\thesection}{\arabic{section}}
\renewcommand{\thesubsection}{\arabic{section}.\arabic{subsection}}
\renewcommand{\thesubsubsection}{\arabic{section}.\arabic{subsection}.\arabic{subsubsection}}

\title{\LARGE \bf
Tall Towers on the Moon
}

\author
{
Sephora Ruppert
$^{a,*}$, Amia Ross$^{a}$, Joost J. Vlassak$^{b}$ and Martin Elvis$^{c}$
\\ \\\small{$^{a}$Harvard College, Harvard University,
Cambridge, MA, USA}
\\\small{$^{b}$Harvard School of Engineering and Applied Sciences, Harvard University, Cambridge, MA, USA}
\\\small{$^{c}$Harvard-Smithsonian Center for Astrophysics, Harvard University, Cambridge, MA, USA}% <-this % stops a space
\thanks{}% <-this % stops a space
\thanks{*Corresponding author}
\thanks{\textit{Email address}: sephoraruppert@college.harvard.edu (S. Ruppert)}
}
\begin{document}

\maketitle
\thispagestyle{plain}
\pagestyle{plain}

\begin{abstract}

The lunar South pole likely contains significant amounts of water in the permanently shadowed craters there. Extracting this water for life support at a lunar base or to make rocket fuel would take large amounts of power, possibly of order Gigawatts. A natural place to obtain this power are the regions of high illumination sometimes called the “Peaks of Eternal Light”, that lie a few kilometers away on the crater rims and ridges above the permanently shadowed craters. The area of high illumination is small at elevations of a few meters. In order to provide sufficient power to support a human base or to mine water much taller solar power towers will be needed.  The limits to how  tall a tower can be built to support the photovoltaic panels has not been investigated. The low gravity, lack of atmosphere, and quiet seismic environment of the Moon all suggest that towers could be built much taller than on Earth. Here we look at the theoretical limits to building tall concrete towers on the Moon. We choose concrete as a first material to investigate because the capital cost of transporting large masses of steel or carbon fiber to the Moon is presently so expensive that profitable operation of a power plant is unlikely. Concrete instead can be manufactured in situ from the lunar regolith. We find that, setting a  minimum wall thickness of 20 cm, towers up to several kilometers tall are stable. The mass of concrete needed, however, grows rapidly with height, from $\sim$ 760 mt at 1 km to $\sim$ 4,100 mt at 2 km to $\mathbf{\sim 10^5}$ mt at 7 km and $\mathbf{\sim 10^6}$ mt at 17 km. We conclude that more detailed investigations of tall lunar towers are warranted.

\end{abstract}

%%%%%%%%%%%%%%%%%%%%%%%%%%%%%%%%%%%%%%%%%%%%%%%%%%%%%%%%%%%%%%%%%%%%%%%%%%%%%%%%
\section{INTRODUCTION}

The South pole of the Moon appears to harbor significant resources in the form of water and organic volatiles in the permanently shadowed regions \cite{Lawrence}. There is considerable interest in harnessing these resources to support a lunar base \cite{Halbach} or to manufacture rocket fuel to resupply rockets at lower cost than bringing the fuel up from Earth \cite{Kornuta}. However, extracting these resources is a power-intensive operation. Kornuta et al. estimate that extraction of 2450 tons/year of water from the permanently dark craters would require power at a level of 0.4 - 1.4 GW (their figure 17) \cite{Kornuta}. 

A promising solution is the nearly continuous energy supply that is potentially available a few kilometers away \cite{Pieters} on high illumination areas, sometimes called the “Peaks of Eternal Light” \cite{Flammarion}. The “Peaks'' are exposed to sunlight for over 90\% of the lunar cycle \cite{Bussey}. However, the illuminated area is only a few square kilometers and much of that area could well become shadowed by other solar towers as lunar development progresses \cite{Ross}, limiting the available power. One way of increasing the potential power output is to build higher. The resulting added power is not just due to an increase in the area provided by tall towers; the illumination is also more continuous as the tower rises above local topography \cite{Glaeser}. Ross et al. showed that for towers up to 20 m tall, the maximum power attainable was of order a few megawatts; instead, for towers from 0.5 - 2 km tall several Gigawatts are achievable \cite{Ross} . Given that Kornuta et al. (figure 17) estimate that extraction of 2450 tons/year of water from the permanently dark craters would require power at a level of 0.4 - 1.4 GW, a need for towers in the kilometer-high range is indicated \cite{Kornuta}. For scale, the Eiffel Tower is 330 m tall \cite{Eiffel} and the tallest building on Earth, the Burj Khalifa in Dubai, is 829 m tall \cite{Baker}. Evidently building comparably tall lunar towers is a challenge, especially as no structures at all have yet been built on the Moon. However, 1/6 gravity on the Moon \cite{Allen}, combined with the lack of an atmosphere and so of winds, and the minimal - though poorly measured -  levels of seismic activity (10$^{10}$ - 10$^{14}$ J/yr) \cite{Vaniman}, suggest that kilometer-scale lunar towers may not be  ruled out. 

Here, we explore the theoretical limits to how tall Moon-based solar towers could be using simple modeling. Determining the tallest structure that can be built with a given material is a field with a history stretching back to Greenhill (1881) \cite{Greenhill}. Since general solutions are hard to find, modeling has to make simplifying assumptions \cite{Wang}. In our calculations we considered limits imposed by both compressive strength of the material and by buckling. In this preliminary study, we focused on towers made of concrete. 

Concrete was studied because transporting materials to the Moon is currently very expensive, of order \$0.5 million/kilogram \cite{Astrobotics}. This makes for an enormous capital cost for a kilometer scale tower, on the order of billions. At these prices lunar water mining would be hard to make into a profitable industry. Instead it has been shown that concrete can be made out of the loose lunar surface material (“regolith”) without using water \cite{Happel}\cite{Khitab}. Doing so would greatly reduce the up-front capital cost as only the relatively lightweight photovoltaic panels would need to be supplied from Earth. Hence, we explore the possibilities for concrete towers on the Moon in this paper.

The following work is far from a complete technical project proposal. Instead, it aims to provide a first look at whether kilometer-scale concrete solar towers are at all promising as an option to exploit the solar capabilities of the “Peaks of Eternal Light” and so worth exploring further.

We used an analytic approach to estimating the stresses in the modeled towers. In this way we could expose the scalings of maximum tower height to the model parameters. We first describe the model in section \ref{theory}. In section \ref{Results} we then present the results after optimizing tower geometry and imposing a minimum wall thickness. We discuss the limitations of these calculations, and so the need for further work, in section \ref{discussion}. We present a summary and our conclusions in section \ref{conclusion}.

\section{THEORY}
\label{theory}

To explore the structural limitations of a concrete tower, we modeled a circular structure that gets exponentially narrower with height. The cross-sectional area at a given height $x$ above the base is described by
\begin{equation}
    A(x)=A_0e^{-kx},
    \label{1}
\end{equation}

where $A_0$ is the cross-sectional area at the base of the tower, $k$ is the exponent by which the tower cross-section shrinks ($k\geq 0$), and $x$ is the height above the base. 
The thickness of the tower’s walls also decreases with the same exponent, $k$. The cross-section of the concrete walls by height is given by

\begin{equation}
    A_c(x)=A_{c,0}e^{-kx},
    \label{2}
\end{equation}
where
$A_{c,0}$ is the cross-sectional area of the walls at the base of the tower. Furthermore,
\begin{equation}
    A_{c,0}=A_0(1-b),
    \label{3}
\end{equation}
where $b$ is unitless and determines the fraction of the tower that is hollow.

\subsection{Stress}

At any point, the tower’s walls are under compressive stress by the weight of the concrete above the point. Because of the circular symmetry of the model, any point of equal height, i.e. any point of a given cross-section, essentially experiences the same amount of stress. As a function of height, the stress is  therefore 
\begin{equation}
    \sigma(x)=\frac{F(x)}{A_c(x)},
    \label{4}
\end{equation}

where $F(x)$ is the weight of the tower section above acting on the cross-section.
\begin{equation}
    F(x)=ma = \rho g \int_x^LA_c(x) {\mathrm  {\mathrm  {d}}}x,
    \label{5}
\end{equation}

where $\rho$ is the density of concrete, 
$g$ describes the gravity on the surface of the Moon, 
and
$L$ is the total height of the tower.
\newline

Applying a safety factor $f_s$ to the load, the resulting stress in the tower is
\begin{equation}
    \sigma(x)= \frac{f_s \rho g}{k} \left(1-e^{k\left(x-L\right)}\right).
    \label{6}
\end{equation}

For an infinitely tall tower, this reduces to $\sigma(x)=f_s \rho g/{k}$, which makes the compressive stress independent of height $x$, i.e. constant throughout the tower. The parameter $k$ can be picked to fix the compressive stress in the structure and optimize the tower's dimensions. It is also worth mentioning that the stress is independent of the base area.
\begin{figure}[]
    \centering
    \includegraphics[scale=0.24]{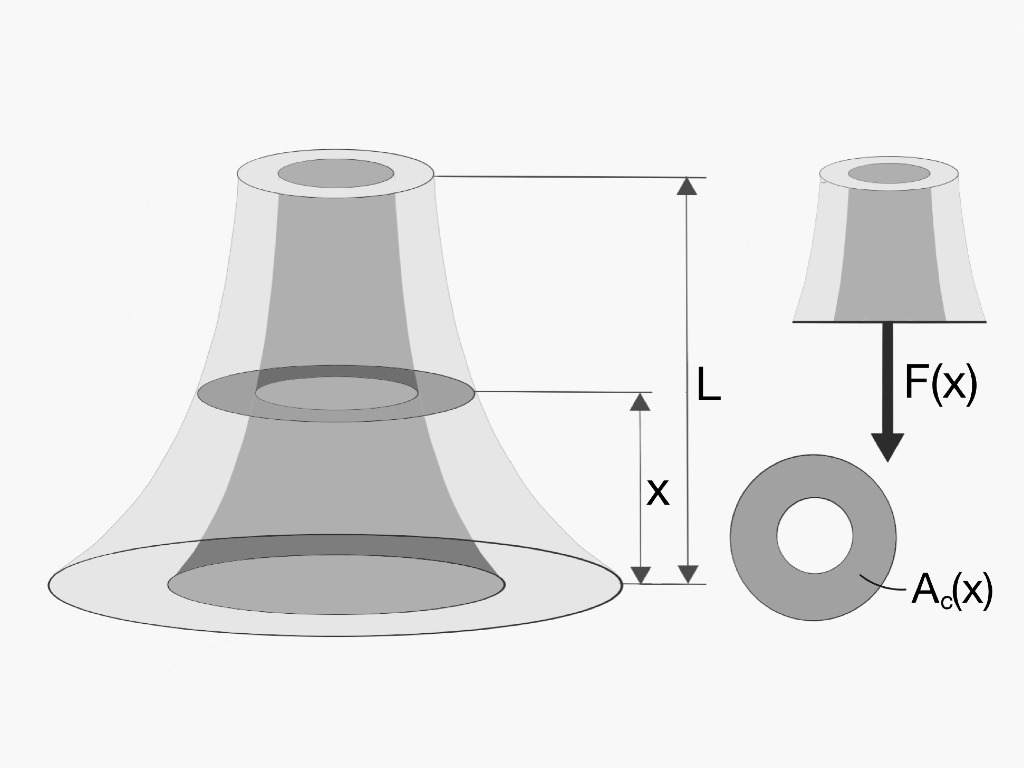}
    \caption{Tower specifications: $L$ is the tower's total height, $x$ the height of a considered cross section, and $A_c(x)$ the cross-sectional area of the concrete at height $x$. $F(x)$ is the total force applied on a given cross section of height $x$ by the weight of the above concrete.}
    \label{TowerSpecifications}
\end{figure}

\subsection{Buckling}

Buckling is the sudden change in shape of a structure under load. For columns that means bending or bowing under a compressive load. If an applied load reaches the so-called critical load, the column comes to be in a state of unstable equilibrium - the slightest lateral force will cause the column to suddenly bend, which decreases the carrying capacity significantly and likely causes the column to collapse. 

The tower in our model is essentially a column and as such, buckles under its own weight at a certain height, also known as self-buckling. To find this critical height, we need to derive the stability conditions for the tower's specific geometry.

For towers with uniform cross section, that is towers that do not get narrower with height ($k=0$), Greenhill \cite{Greenhill} found that the critical self-buckling height is
 \begin{equation}
   L_{c}\approx \left(7.8373\,{\frac {EI}{\rho gA_c}}\right)^{1/3}
   \label{7}
\end{equation}

where $E$ is the elastic modulus, $I$ is the second moment of area of the beam cross section, $\rho$ is the density of the material, $g$ is the acceleration due to gravity and $A_c$ is the cross-sectional area of the body \cite{Greenhill}.

\subsubsection*{Self-Buckling of a column of non-uniform cross-section (k$>$0)}

The Euler–Bernoulli theory, also known as the classical beam theory, provides means of calculating the deflection behaviour of beams. The theory entails the bending equation, which relates the bending moment of a beam or column to its deflection:
\begin{equation}
    {\displaystyle M(x)=-EI{\mathrm{d}^{2}y \over \mathrm{d}x^{2}}}
    \label{8}
\end{equation}
where 
$M(x)$ is the bending moment at some position $x$,

$E$ is the elastic modulus,

$I(x)$ is the second moment of area of the beam's cross-section at $x$,

$y(x)$ describes the deflection of the beam in the y-direction at $x$.
\newline

\begin{figure}[]
    \centering
    \includegraphics[scale=0.2]{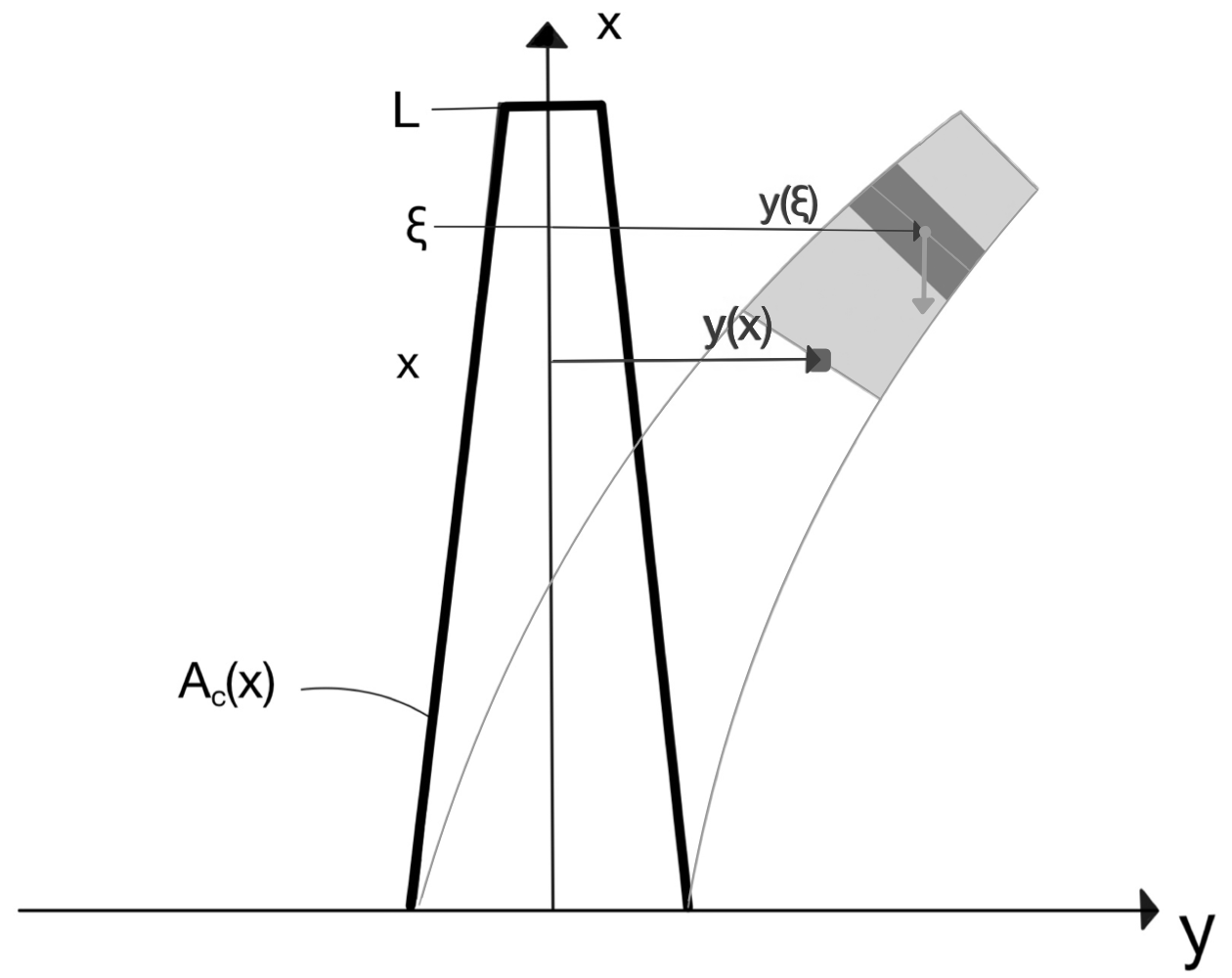}
    \caption{Quantities and variables of the buckling model: $L$ is the tower's total height, $x$ the height of a considered cross section, and $A_c(x)$ the cross-sectional area of the concrete at height $x$. $\xi$ is the height of an infinitesimally thin sliver of concrete, whose weight acts on a given cross section at height $x$. $y(x)$ and $y(\xi)$ quantify the tower's horizontal displacement from the $x$-axis at heights $x$ and $\xi$, respectively.}
    \label{bucklingModel}
\end{figure}

For this specific model (see figure \ref{bucklingModel}), we can define some useful quantities:
The linear weight density of the column is given by $w(x)=A_c(x)\rho g$, where $A_c(x)=A_{c,0}e^{-kx}$ is the cross-sectional area of the concrete at any given height. The second moment of inertia is $I(x)=\int_Ay^2dA=A_0^2\left( 1-b^2\right)e^{-2kx}/4\pi$.
\newline

We define $\xi$ to be the height above the base of a elementary mass weighing on the horizontal plane of interest at height $x$. The moment at height $x$ can be written as
\begin{equation}
    M(x) = \int_x^L(w(\xi) d\xi)(y(\xi)-y(x))
    \label{9}
\end{equation}
Substituting equation \ref{9} into the bending equation, equation \ref{8}, gives
\begin{equation}
    EI\frac{{\mathrm  {\mathrm  {d}}}^2y}{{\mathrm  {\mathrm  {d}}}x^2}=\int_x^Lw(\xi)(y(\xi)-y(x))d\xi
    \label{10}
\end{equation}
Substituting the expressions $w(\xi)$, $I(x)$ and $A_c(\xi)$ and simplifying yields
\begin{equation}
    \frac{{\mathrm  {\mathrm  {d}}}^3y}{{\mathrm  {\mathrm  {d}}}x^3}-2\frac{{\mathrm  {\mathrm  {d}}}^2y}{{\mathrm  {\mathrm  {d}}}x^2}-\alpha\left(e^{2kx-kL}-e^{kx}\right)\frac{{\mathrm  {\mathrm  {d}}}y}{{\mathrm  {\mathrm  {d}}}x}=0
    \label{11}
\end{equation}
where the constant
$\alpha={4\pi\rho g}/{(b+1)EA_0k}$.

Setting ${{\mathrm  {\mathrm  {d}}}y}/{{\mathrm  {\mathrm  {d}}}x}=\eta$ and $\gamma=kx$ results in the following ordinary differential equation:
\begin{equation}
    \frac{{\mathrm  {\mathrm  {d}}}^2\eta}{\mathrm  {d}\gamma^2}-2\frac{d\eta}{d\gamma}-\beta\left(e^{2\gamma-\lambda}-e^{\gamma}\right)\eta =0
    \label{12}
\end{equation}

where
$\lambda=kL$ and $\beta={\alpha}/{k^2}={4\pi\rho g}/{(b+1)EA_0k^3}$.
\newline
Since the tower is fixed against deflection (clamped column end) at its base and is unconstrained and therefore unbent at the top (free column end), we have the following boundary conditions:
$\eta(0)=0$ (clamped),    $\eta'(L)=0$ (free end).
\newline

Buckling will occur when equation \ref{12} has a non-trivial solution. This requirement yields a critical $\lambda$ (or $L$) for a given $\beta$, at which the tower will buckle and which can be calculated numerically. Applying a safety factor $f_b$ to the loads makes the normalized length $\lambda=kL$ and the normalized load $\beta={4\pi f_b \rho g}/{(b+1)EA_0k^3}$.

\section{Results}
\label{Results}
\subsection{Safety Factors}
The model calculates an absolute maximum height for a tower before failure. In any realistic tower a safety factor (S.F.) is needed. The disturbances on the Moon are presently much lower than on Earth, but the vibrations created by the mining activity that these towers would support would make for additional stresses. The possibility of vehicle collisions with the towers, e.g. during maintenance operations, must also be considered.

For concrete structures S.F. of 1.2 are commonly applied to compressive loads \cite{Hassoun}. Structures at risk of buckling usually require much higher additional safety factors. Since the exact building environment is difficult to predict at this time and the construction would be an costly endeavour, more conservative S.F. between 3 and 4 are likely warranted \cite{Rutheravan}. 

At this time, it is difficult to say which exact S.F. would be appropriate, as there are no norms or examples of structural engineering on extraterrestrial bodies. The high cost of transporting building essentials to the Moon, might lead to the use of lower S.F. to save material. On the other hand, it could also be a reason to raise the S.F. to guarantee the structure’s longevity. Until more details on future transport costs to the Moon and lunar manufacturing costs become clearer, it is hard to predict an appropriate  safety factor.
\newline

Throughout the paper the safety factors for compressive stress and buckling are denoted by $f_s$ and $f_b$, respectively. In our analysis, we present results for  $f_s$  and $f_b = 1$ to obtain the limiting case geometry of the tower; building a tower any taller or otherwise differently shaped, could result in immediate failure.  Once a reasonable S.F. is determined, those results can be rescaled appropriately. Varying $f_b$  changes the $k$-value proportionally (see equation \ref{13}), which decreases the height limit based on wall thickness (section \ref{ResultsWallthickness}). The effect of the safety factor on the buckling height is discussed in section \ref{ResultsBuckling}.

The minimum thickness of the tower walls is determined by both the safety factors and the exact material characteristics of lunar concrete, neither of which are completely determined at this time. Throughout this paper a minimum thickness of 20 cm is used. This value already includes extensive safety factors as it is a building guideline intended for load bearing exterior walls in tsunami and earthquake-prone environments (see e.g. the Caribbean Disaster Mitigation Project \cite{Disastermitigationproject}).

\subsection{Failure due to compressive stress}
\label{ResultsStress}

”Concrete” describes a range of material with compressive strengths ranging from under 10 MPa to over 100 MPa \cite{Argolu}. Lunar “concretes” are unlikely to use water and so would stretch the definition of the term. A sulfur based concrete that can be made out of lunar regolith has a compressive strength of about 30 MPa \cite{Omar}. This is then a reasonable fiducial value to use in our calculations. 

In our model, the stress throughout the tower is constant, given a fixed density and k-value (see equation \ref{6}). To take full advantage of the compressive strength of the concrete, we can set $\sigma_{max}=30$ in equation \ref{6} to obtain the $k$-value:
\begin{equation}
    k= \frac{f_s \rho g}{\sigma_{max}}
    \label{13}
\end{equation}
where $\rho=2400$ $kg$ $m^{-3}$,

$g=1.62$ $m$ $s^{-2}$,

$\sigma_{max}=30$ MPa, 

$f_s$ is the safety factor applied to the load.
\newline

Literature offers several possible densities for lunar concrete ranging from 2200 kg m$^{-3\,}$ \cite{Omar} to 2600 kg m$^{-3}$ \cite{Happel}. We are therefore using a density of 2400 kg m$^{-3}$, which is the same as that of typical terrestrial concretes \cite{Jones}. 

For $f_s=1$ we find $k=0.00013$ $m^{-1}$. In the case of a tower on the scale of a lunar radius, this value is an underestimate, as acceleration due to gravity ($g$) decreases with height.
Even for shorter towers of height $L$, the stresses will not be exactly uniform at 30 MPa, but instead will decrease by $\Delta \sigma=\left(f_s \rho g /k \right)\exp[k(x-L)]$ at any height $x$ (from equation \ref{6}). For a 100 m tower, $\Delta \sigma$ constitutes a $97\%$ change in compressive stress at the base, i.e. just above the base the compressive stress is only $3\%$ of the maximum value (30 MPa). This  change decreases exponentially as the tower height increases (e.g. $>89\%$ for 1 km, $>27\%$ for 10 km, $>0.0002\%$ for 100 km).
For any tower height, the compressive stress is below 30 MPa everywhere and the tower is stable against compression.
\newline

Theoretically, a tower with no additional forces acting upon it is only limited in height by the stress capacity of the material. In practice, the walls of the modeled tower will ultimately become too thin, to support any secondary structures such as solar panels (see section \ref{ResultsWallthickness}).

Additionally, horizontal forces caused by impacts or vibrations cannot be ruled out. Because of this, we need to consider the risk of buckling and adjust the critical height accordingly (see section \ref{ResultsBuckling}).

\subsection{Wall thickness}\label{ResultsWallthickness}
The stress in the tower is independent of the wall thickness at the base area and stays (roughly) constant as the walls become thinner with height (see equation \ref{6}). Theoretically, then, the walls could be infinitely thin, and the tower would still be self-supporting. 
Realistically, however, the tower’s walls should always exceed a minimum thickness (here 20 cm) \cite{Disastermitigationproject}.
\newline 

In our model the wall thickness is indirectly defined through the cross-sectional area of  the  walls, defined in equation 2 by 
$$A_c(x) = (1-b)A_0e^{-kx},$$
where $x$ is the height above the base,

$A_0$ is the base area,

$k = 0.00013$ $m^{-1}$ is the factor of decay optimized for our model,

$b$ is a real, positive number so that $b < 1$ and is the fraction of the tower cross-section that is hollow. Larger values of $b$ correspond to thinner walls.
\newline

How high a given tower can be while still exceeding a minimum wall thickness depends on the $b$-value and the base area $A_0$.

\begin{figure}[]
    \centering
    \includegraphics[scale=0.5]{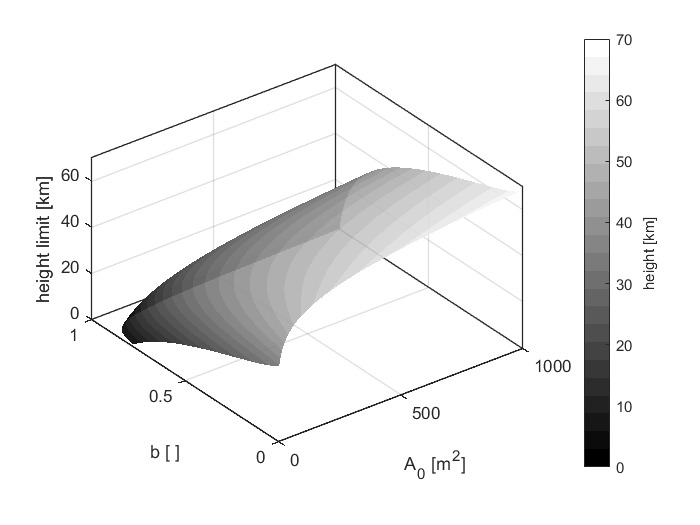}
    \caption{Tower height limit based on wall thickness $L$ (km) versus $b$ ( ) and base area $A_0$ (m$^2$). $k=0.00013$ $m^{-1}$.}
    \label{wallThickness1}
\end{figure}

\begin{figure}[]
    \centering
    \includegraphics[scale=0.52]{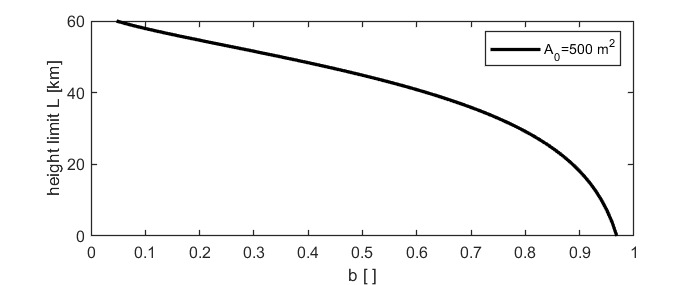}
    \caption{Tower height limit based on wall thickness $L$ (km) versus $b$ ( ) for a base area $A_0$ of 500 m$^2$. $k=0.00013$ $m^{-1}$.}
    \label{wallThickness2}
\end{figure}

\begin{figure}[]
    \centering
    \includegraphics[scale=0.52]{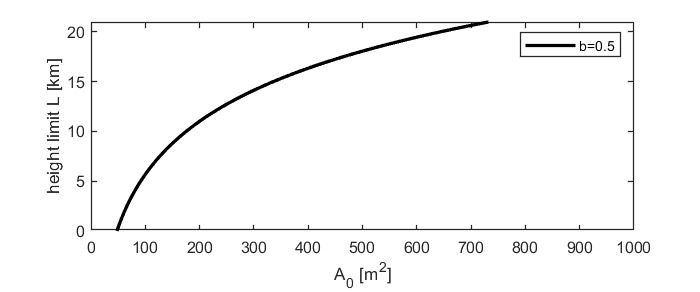}
    \caption{Tower height limit  based on wall thickness $L$ (km) versus base area $A_0$ (m$^2$) for a $b$-value of 0.5. $k=0.00013$ $m^{-1}$.}
    \label{wallThickness3}
\end{figure}

To demonstrate the resulting trends, figure \ref{wallThickness1} shows the limiting tower height as a function of $b$ and base area. Figures \ref{wallThickness2} and \ref{wallThickness3} show cross-sections of figure \ref{wallThickness1} for a 500 m$^2$ base area and a 0.5 $b$-value.

\subsection{Buckling}\label{ResultsBuckling}
 If the buckling load of a structure is exceeded, any imperfection or perturbation, no matter how small, causes the building to buckle \cite{Beer}. (Buckling might occur at lower loads for large disturbances; safety factors take this into consideration.)

\begin{figure}[]
    \centering
    \includegraphics[scale=0.42]{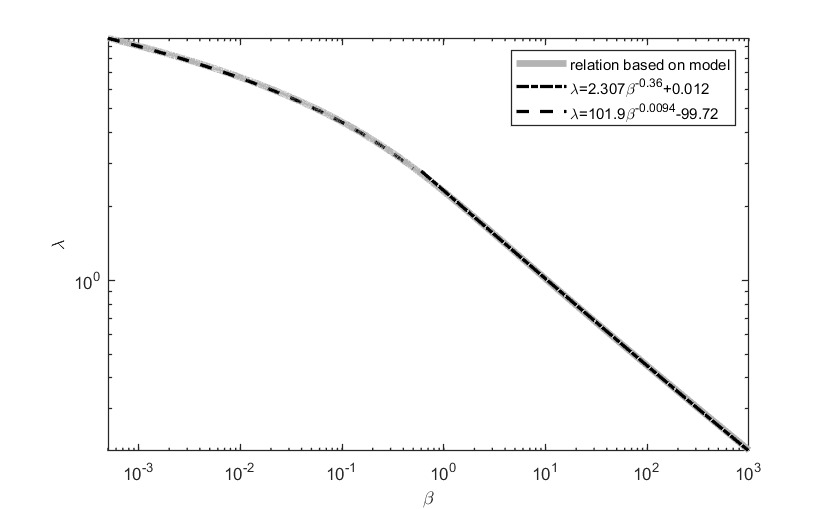}
    \caption{Normalized length, $\lambda$ versus normalized load, $\beta$, on logarithmic scales. The yellow line shows the $\lambda \sim \beta^{-0.0094}$ power law fit for the range $\beta=0.05-0.5$. The red line shows the $\lambda \sim \beta^{-0.36}$ power law fit for the range $\beta=0.5-1000$.}
    \label{lambdaVSbeta}
\end{figure}

The boundary value problem describing the tower's buckling behavior is given  by equation \ref{12} and the corresponding boundary conditions outlined in section \ref{theory} - one clamped and one free. To simplify the problem it is convenient to use the normalized length $\lambda  = kL$, and the normalized load $\beta =4\pi f_b\rho g/\left(\left(b+1\right)EA_0k^3\right)$. A numerical solution yields the $\beta$-$\lambda$ values shown in figure \ref{lambdaVSbeta}.

The analysis uses $g = 1.62$ $m$ $s^{-2}$ as the lunar acceleration due to gravity. This is accurate at all the heights considered here, as the maximum values  are much smaller than the Moon’s radius (1737.1 km \cite{Allen}). The density of concrete used here is 2400 kg m$^{-3}$ \cite{Jones}.

\begin{figure}[]
    \centering
    \includegraphics[scale=0.5
]{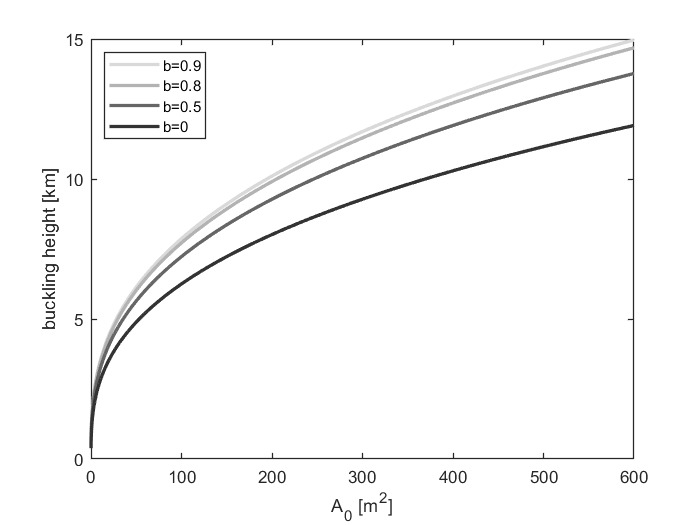}
    \caption{ Critical buckling height, $L_c$ (km) as a function of tower base area $A$ ($m^2$) for $k=0.00013$ $m^{-1}$ and various values of $b$. A safety factor of $f_b = 1$ is used here.}
    \label{bucklingheight}
\end{figure}

Figure \ref{bucklingheight} shows the tower's critical buckling height as a function of the base area for $k=0.00013$ $m^{-1}$ and the safety factor $f_b = 1$. Tower heights of order tens of kilometers are achievable.

To assure safe results, an additional safety factor $f_b$ can be considered. The normalized load, $\beta$,  is proportional to the safety factor and will therefore increase linearly with it. As a result the critical buckling height of the tower decreases. The new theoretical buckling height, will be able to support $f_b$ times the actual load. 

The factor by which the theoretical buckling height decreases depends on the relationship between $\lambda$ and $\beta$ for a given setup. Over a wide range ($\beta$ = 1 - 1000) this relation is well modeled by a power-law relation: $\lambda=2.307\beta^{-0.36}+0.012$. (See figure \ref{lambdaVSbeta}.) The maximum tower height scales down by this $f_b^{-0.36}$ factor for $\beta$ between 0.5 and 1000, which roughly correlates to heights under 20 km for $k=0.00013$ $m^{-1}$. For $\beta$ between 0.05 and 0.5, the maximum tower heights scale down by a factor of $f_b^{-0.0094}$ which roughly corresponds to heights above 20 km for $k=0.00013$ $m^{-1}$. 

\subsection{Optimizing the maximum height}

The ideal tower should be both tall and require as little concrete for construction as possible. To keep the mass of building material low, the tower’s walls should be as thin as possible, that is, parameter $b$ should be maximized. As the $b$-value increases, however, the maximum height decreases (see figure \ref{wallThickness2}). The buckling height, on the other hand, increases with $b$ (see section \ref{ResultsBuckling}). This suggests a trade-off between maximizing the buckling height and  maximizing the height limit based on wall thickness. The maximum tower height will be then be the smaller of the two heights. To optimize maximum height, $b$ must be picked carefully. 

\begin{figure}[]
    \centering
    \includegraphics[scale=0.38]{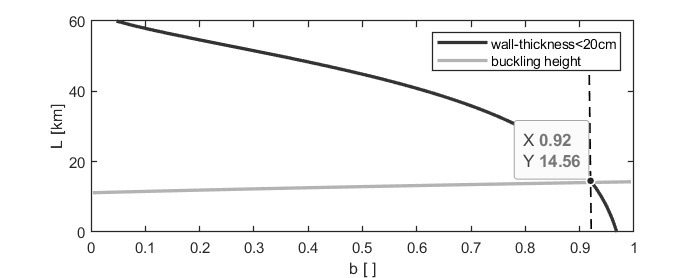}
    \caption{Height at which the wall thickness reaches 20 cm (km) and buckling height $L_c$ (km) versus $b$ for $f_b = 1$ and $A_0=500$ $m^2$. The vertical dashed line at $b$ = 0.92 marks the intersection.}
    \label{intersect}
\end{figure}

Figure \ref{intersect} shows both the height at which the minimum wall thickness is reached and the buckling height for a given $b$-value. The point at which the two curves cross, at $b\sim 0.92$ is where the dominant limiting factor changes. For $b< 0.92$, buckling dominates, for $b > 0.92$ wall thickness dominates. 

\begin{figure}[]
    \centering
    \includegraphics[scale=0.5]{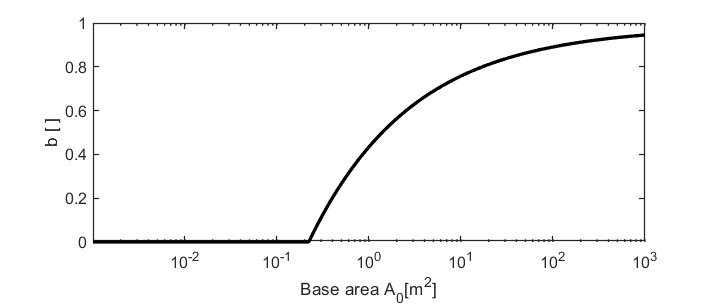}
    \caption{Ideal $b$ ( ) to maximize height versus base area $A_0$ (m$^2$) for $f_b = 1$.}
    \label{idealAvb}
\end{figure}

For a base area of 500 m$^2$, 0.92 is the ideal $b$-value. Similarly, $b$ can be found for other base areas. Figure \ref{idealAvb} shows the relationship between the base area and the ideal $b$, so that the total maximum height is as great as possible. For values of $A_0\textgreater 10$ m$^2$ and for $A_0\textgreater 100$ m$^2$ the optimum value of $b$ is $\textgreater 0.8$ and $\textgreater 0.9$, respectively. The optimal $b$ value approaches 1, as the base area increases. This relationship will be kept in mind when choosing $b$.

\subsection{Mass of maximum height tower}

Given the model parameters we can calculate the mass of the concrete required to build a tower of height $L$. 
\begin{equation}
    M=\rho \int_0^L A_0(1-b)e^{-kx}dx=\frac{\rho A_0}{k}(1-b)(1-e^{-kL}).
    \label{14}
\end{equation}
where $b$ is the wall thickness parameter chosen in relation to $A_0$ based on figure \ref{idealAvb}.

The mass of interest is for a tower at the buckling height, with the smallest possible base area for a given height and the thinnest possible walls. Based on figures figures \ref{bucklingheight} and \ref{idealAvb}, the base area $A_0$ and the wall thickness parameter $b$ are optimized for each tower height $L$. These parameters give the minimum concrete mass requirement for realistic tower proportions.

\begin{figure}[]
    \centering
    \includegraphics[scale=0.5]{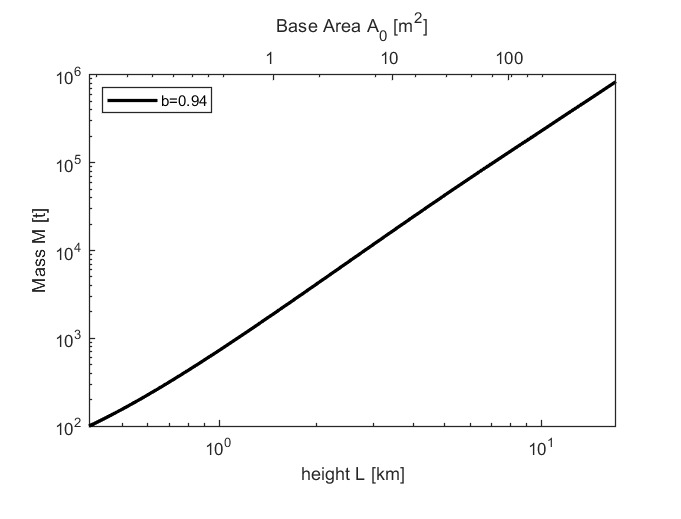}
    %\vspace{-1.5cm}
    \caption{Mass of concrete $M$ (mt) needed to build a tower of a given height $L$ (km) and base area $A_0$ (m$^2$), for $f_b = 1$. For every height-base area pair, the ideal $b$-value from figure \ref{idealAvb} is used. Note that the scale for $A_0$ is not linear, but chosen to reflect the relationship between height $L$ and base area $A_0$. Also note that the scales for $M$ and $L$ are logarithmic; the mass grows rapidly with height.}
    \label{Mass}
\end{figure}

Figure \ref{Mass} shows the mass of concrete required against both the total height of the tower and the base area required. Note that the horizontal axis with the values for the base area is not linear. Instead the scale is chosen so that a given height matches up with the ideal base area. 

\section{Discussion}
\label{discussion}
\subsection{Optimal tower height}
\subsubsection{Height Limit due to Wall Thickness}

It is important for any freestanding structure to support its own weight. For a tower made of lunar concrete that means that the compressive stress must not exceed 30 MPa anywhere \cite{Omar}. In an equal stress structure, as modeled here, the stress state at all points of the body is the same. This gives the most efficient use of building material.
\newline

Because of this our model is arranged so that an infinitely tall concrete tower under the Moon’s surface gravity is an equal stress structure at its stress capacity. Here, the cross-sectional area of the tower’s walls decreases exponentially with height by a factor of
$k = 0.00013$ $m^{-1}$. For finite heights, the stress distribution
is no longer perfectly uniform. However, it is always below 30 MPa, allowing the tower to still be self-supporting. 

$k = 0.00013$ $m^{-1}$ is chosen for a hollow concrete structure. If the tower were more complex, i.e., interior structures such as floors were added or multiple building materials included, the ideal $k$-value would have to change to reflect that. Changes to the $k$-value of order $10^{-4}$ (i.e. factors of order 2), do not change the maximum heights significantly. 
\newline

The tower’s maximum stress is independent of its cross-sectional area, as well as that of its walls. Theoretically, the wall could therefore be infinitely thin, and the tower would still be self-supporting. However, concrete is an aggregate material with a range of particle sizes that do not allow arbitrarily thin walls. This property sets a minimum practical concrete wall thickness.
\newline

According to the Caribbean Disaster Mitigation Project, a load bearing exterior wall should be a minimum of 20 cm thick \cite{Disastermitigationproject}. This value already includes extensive safety factors as it is a building guideline intended for government buildings in tsunami- and earthquake-prone environments.

Adopting a thickness minimum sets a limit to the tower height, as the tower walls become thinner with height, but may not fall short of the minimum wall thickness. The maximum height based on wall thickness increases with the base area $A_0$ and decreases as the hollow fraction $b$ of the tower’s cross section increases (see figure \ref{wallThickness1}).
\newline

\subsubsection{Height Limit due to Buckling}
Next to compressive behaviour, it is important to consider their buckling behaviour. The tower’s buckling behaviour for a fixed $k$-value is dependent on the cross-sectional area of its base ($A_0$) and the relative thickness of the walls ($b$). 

In our model the critical height due to buckling can be infinite, given a sufficiently big base area. The highly illuminated regions on the surface of the Moon offer limited construction area, though \cite{Ross}. This limits how big the base area and therefore the critical height can be.
\newline

From figure \ref{bucklingheight}, we know that a tower with a thinning rate $k=0.00013$ $m^{-1}$ with a 500 m$^2$ base area has a buckling height between 12 km and 15 km depending on the wall thickness parameter $b$. A solid tower has the lowest buckling height at 12 km. The buckling height increases as the walls get thinner.

As the wall thickness decreases, the buckling height increases. Therefore, $b$ should be made as large as possible to keep the wall thickness low. The walls cannot be arbitrarily thin, though. In this analysis we required that they always exceed 20 cm. In this model, the walls become exponentially thinner with height and will therefore always fall short of the minimum thickness at some height. The height at which this happens decreases as $b$ increases. The buckling height has the reversed relationship with the wall thickness and increases with $b$.
\newline

\subsubsection{Combined Buckling and Wall Thickness Constraints}
For the final tower to be as high as possible, $b$ must be chosen so that the height at which the minimum wall thickness is reached and the buckling height are equal. This is dependent on the base area. The ideal $b$-value increases with the base area and approaches 1. For base areas between 10 m$^2$ and 1000 m$^2$, the $b$-value falls in between 0.75 and 0.95. (See figure \ref{idealAvb}.)
\newline

A tower with a 500 m$^2$ base, $k=0.00013$ $m^{-1}$ and $b=0.92$ would reach its maximum height at 14 km. Such a tower would require 520 thousand tons of concrete see figure \ref{Mass}). 

Figure \ref{Mass} shows that a 1 km tall tower of a 1 m$^2$ base needs $\sim$ 760 mt of concrete, while a tower of 2 km height and 2 m$^2$ base requires a mass of concrete of around 4,100 mt. By Earth standards, these seem like remarkably small base areas for such tall towers and we would expect them to be unstable against external forces. This intuition stems from the strong winds and seismic activity that make it difficult to build tall towers on Earth. However, there are no external forces comparable to terrestrial winds on the Moon, allowing for more slender towers. In practice, especially near to a human base or to mining operations, there will be other forces acting. Estimating the strength of these, and so the extra base area and/or thickness required, will need to be studied.

The required mass of the towers grows rapidly with height; by 7 km (70 m$^2$ base area) the mass is 10$^5$ mt, and by 17 km (700 m$^2$ base area) has almost reached 10$^6$ mt. 
\newline

Figure \ref{Mass} demonstrates that the mass and volume of regolith that needs to be processed into concrete in a reasonable time is quite likely to be the limiting factor for some time. If we require a construction time of 1 year, then building a 2 km tower would require processing 11 mt/day of concrete, about 4.6 m$^3$/day. A 1 km tower would require a factor 5  lower rates. These seem like plausible numbers for a decade or two from now.

\subsection{Weight of  solar panels}
The weight of the solar panels needs to be considered, as the towers have no function without them. At their thinnest, the tower’s walls are 20 cm thick, which amounts to a cross-sectional area of at least 0.126 m$^2$ and a cross-sectional circumference of 128 cm. Since the ratio of the concrete’s cross-sectional area to the circumference is the greatest at this point, this is where the solar panels will have the greatest impact on the load. 

The density of concrete is 2400 kg
 m$^{-3}$ and the mass of a state of the art triple junction  solar panel for use in space is  $\sim$ 2 kg m$^{-2}$ \cite{Spectrolab}. At the tower’s narrowest part the concrete will have a mass of 310 kg m$^{-1}$. Solar panels will add to this load by 0.6\%. We conclude that the weight of the photovoltaic panels is negligible for towers within this model.
 
 \subsection{Future consideration}
 This paper is only intended to provide a first estimate of the height limitations of lunar concrete towers and is not an exhaustive analysis of possible designs and failure modes. As a result there are many open issues to address before implementing solar towers on the Moon, each involving multiple different factors: e.g., material properties, geometry, reinforcements. In this section, we outline some of these considerations that should be explored in further studies.
 \newline
 
 \subsubsection{Shell Buckling}
 To determine the buckling limit, we performed a beam buckling analysis based on the Euler–Bernoulli theory, which is useful in predicting the buckling behaviour of beams, columns, and other members. This formalism, however, neglects imperfections and second-order deformations that can lead to local buckling phenomena in thin-shell structures, i.e. shell buckling. 

A shell is a thin, curved rigid structure, whose thickness is small compared to its other dimensions. Such structures have a significantly lower critical buckling load than the Euler–Bernoulli values \cite{Hogendoorn}. 

Since this paper finds the optimized tower to be $\sim$ 90\% hollow (see section \ref{ResultsBuckling}), shell buckling could be especially relevant \cite{Bushnell} and might decrease the optimal height-to-mass ratio for a given base area by imposing additional height limitations. 

Predicting a tower’s shell buckling behaviour is a complex issue, requiring sophisticated analyses beyond the scope of this paper. This is a topic for future detailed investigation.
\newline

\subsubsection{Lunar concrete}
While some concepts for lunar soil-based building material have been studied using simulated lunar regolith \cite{Happel}\cite{Omar}, true lunar concrete has not yet been manufactured from lunar regolith, so the properties of lunar concrete have not been studied.  There is no guarantee such materials are realizable with the necessary properties from real lunar regolith. 

Further work into the properties of lunar concrete is ongoing. The prospects for using lunar regolith samples are growing \cite{Qian}\cite{NASA}. As these results come in, lunar tower designs can be modified accordingly.
\newline

Once there is a better understanding of the properties of lunar concrete, it will become  important to study the thermal expansion effects on tall structures. Ground level temperature differences at the Moon’s south pole can reach 200 Kelvin \cite{Williams}. Even at high elevations there will be intervals of a few days when sunlight does not reach the towers leading to similar temperature variations. Such large temperature  excursions can have major effects on the mechanical and deformation properties of conventional concrete \cite{Kodur} and so need to be characterized for lunar regolith-derived concrete.
\newline

The potential for erosion through blast ejecta kicked up by landers operating nearby is substantial, given the restricted area near the best solar tower locations. Estimates predict that a 200 ton lunar lander will blow 1,000 tons of ejecta (including fist-sized rocks at 100 km h$^{-1}$), part of which will be blasted over 20 km away from the landing site \cite{Metzger}. Methods of mitigating  the effects of these ejecta,e.g., with landing pads, will need careful study. By the time construction on the Moon is feasible, there will likely be several prospective lunar-based building materials with a range of differing properties to choose from. Depending on the location of the landing sites and the quality of landing pads, the blast ejecta can have dramatic effects on the longevity of the towers. The resistance of each form of concrete to such erosion needs to be considered next to mass and material strengths when deciding on a building material.
\newline

\subsubsection{Transport and Infrastructure}
Although much work is being done on the topic, we do not know yet what lunar transportation and infrastructure on the Moon will be available when tall solar towers are constructed. The actual limitations of lunar construction may lie in other factors than the strength of the towers. These limitations include: material and labor cost and availability, safety factors, mechanical limitations (e.g. rotating solar panels), and height limitations to avoid flight risks.

\subsection{Alternative uses for tall lunar towers}
The obvious use of concrete towers to increase the solar power that can be captured may not be the only use for tall lunar towers. Towers anywhere on the Moon’s surface have been discussed to serve alternative purposes. For example such tall towers can:
\begin{itemize}
  \item provide ambient lighting to the permanently shadowed regions near the “Peaks of Eternal Light” to create suitable diurnal rhythms, if diffuser "sails" are installed atop the towers \cite{Schrunk}. These could be more efficient than conversion of photovoltaic electricity to microwave or laser transmission at $\sim$50\% \cite{Brown}.
  \item enable longer distance line-of-sight communication, power transmission using microwaves or lasers, and navigation. The horizon at human height (1.7 m) is only $\sim$2.4 km away \cite{Plait} on the Moon. (c.f. $\sim$4.7 km on Earth). For a 1 km tower the horizon is $\sim$59 km distant. To create a continuous communication network it requires just eleven 1 km towers per hundred thousand square kilometers.
\end{itemize}

\section{CONCLUSION}
\label{conclusion}
The amount of solar power that can be generated from the high illumination regions near the lunar poles (the “Peaks of Eternal Light”) depends on how tall the arrays of photovoltaic panels can be constructed.
\newline

As a first step toward determining these limits we have studied the stability of concrete towers on the Moon against compressive failure and buckling, for heights from tens of meters to tens of kilometers and have also estimated the mass, and required production rate of concrete needed to build them. Concrete was selected as there are concepts for making “concrete” from the lunar regolith, which saves the cost of importing construction material from Earth. We modeled circular towers with diameters that become exponentially narrower and wall thicknesses and thinner (with exponent $k$) with height until a minimum wall thickness is reached. These properties lead to the compressive stress being constant throughout the tower.
\newline

We find that the compressive stress in the tower is minimized for an exponent $k = 0.00013$ m$^{-1}$. The maximum height is reached for a fraction $b$ of the tower cross-section that is hollow, which increases with the base area and lies in the 0.9 - 1 range for base areas above 100 m$^2$. The base area required to support the tower, and therefore the hollow fraction of the cross-section, increases drastically with height from 0.5 m$^2$ at 1 km to 200 m$^2$ at 10 km.
\newline

We find that kilometer-scale concrete towers on the Moon are  stable against both compressive failure and buckling up to multiple kilometer heights. The solar panels add negligible mass, as their maximum contribution is 0.6\% of the tower mass per meter height (at the towers narrowest section). As more work on lunar concrete and lunar habitats is being done, the tower model proposed by this paper may be adjusted to account for logistical and structural challenges.
\newline

The more likely limiting factor is the mass of concrete needed. To reach 1 km height requires $\sim$ 760 mt. If we require a construction time of 1 year, then 2 mt/day would have to be processed, which does not seem unreasonable for 20 years from now. However the  required mass of concrete grows rapidly with height to 4100 mt at 2 km, due to the rapidly increasing base area.  Local topography may also limit the use of large base areas.
\newline

Future studies should consider metal truss frame towers as they are likely to require much less mass. At sufficiently low transport costs metal trusses imported from Earth may be cheaper than concrete. The trade space between methods should be investigated.
\newline

Other uses for tall towers on the Moon may be worth investigating, for example, as nodes in a communications, navigation and/or power distribution network.

\section*{Acknowledgments}
SR thanks Harvard University’s Program for Research in Science and Engineering (PRISE) for providing support during our research efforts. We thank Robin Wordsworth for helping this project get started. AR thanks the Smithsonian Astrophysical Observatory for support during this project. We thank Hongyan Ma for helpful comments.

\bibliography{refs}
\newpage

\onecolumn
\section*{APPENDIX: Nomenclature}

\begin{table}[h]
\centering
\renewcommand{\arraystretch}{1.5}
\begin{normalsize}
%\resizebox{\columnwidth}{!}{
\begin{tabularx}{\textwidth}{l l X}
\toprule
\textbf{Variable} & \textbf{Units} & \textbf{Description}  \\%[5pt]
\midrule
$A_0$                                            & m$^2$             & Cross-sectional area of the tower’s base                                                                                                             \\%[10pt] 
%\hline
$b$                                             & -           & Fraction of the cross section that is hollow, thereby describing the thickness of the walls through reversed relationship, 0 $\leq$ $b$ \textless{} 1 \\%[20pt] 
%\hline
$f_b$                                            & -              & Safety factor applied to loads in buckling calculations                                                                                              \\%[10pt] 
%\hline
$f_s$                                            & -              & Safety factor applied to loads in stress calculations                                                                                                \\%[10pt] %\hline
$g = 1.62$                     & m s$^{-2}$          & Lunar acceleration due to gravity \cite{Hirt}                                                                                                                 \\%[10pt] 
%\hline
$k$                                             & m$^{-1}$            & Factor of decay describing how the cross-sectional area changes with height                                                                          \\%[10pt] 
%\hline
$L$                                             & m              & Total height of the tower                                                                                                                            \\%[10pt] 
%\hline
$M$                                             & mt             & Mass of of concrete required to build tower                                                                                                          \\%[10pt] 
%\hline
$x$                                             & m              & Height above base                                                                                                                                    \\%[10pt] 
%\hline
$\beta$                                             & -              & $\beta=4\pi\rho g/((b+1)EA_0k^3)$, normalized load used in the buckling analysis                                                                                   \\%[10pt] 
%\hline

$\lambda$                                             & -              & $\lambda = kL$, normalized length used in the buckling analysis                                                                                              \\%[10pt] %\hline
$\rho= 2400$            & kg m$^{-3}$         & Density of concrete (see section \ref{ResultsStress})                                                                                                                                  \\%[10pt]
%\hline
$\sigma$ & Pa             & Compressive stress                                                                                                                                               \\%[10pt]
\bottomrule
\end{tabularx}
%}
\end{normalsize}
\end{table}

\end{document}